\documentclass[aps,preprint,showkeys,
superscriptaddress,
amsmath,amssymb,
aps,
]{revtex4-1}

\usepackage{graphicx}
\usepackage{dcolumn}
\usepackage{bm}
\usepackage{color}
\usepackage[utf8]{inputenc}

\usepackage{mathtools}
\usepackage{empheq}
\usepackage[skins,theorems]{tcolorbox} 
\tcbset{highlight math style={enhanced,
  colframe=black,colback=white,arc=4pt,boxrule=1pt}}

\usepackage{algpseudocode}
\usepackage{algorithm}
\usepackage[english]{babel}

\usepackage{etoolbox}
\usepackage{amsmath}

\newcommand{\algstrut}[1][\algruledefaultfactor]{\vrule width 0pt
depth .25\baselineskip height #1\baselineskip\relax}
\newcommand*{\algrule}[1][\algorithmicindent]{\hspace*{0.5em}\vrule\algstrut
\hspace*{\dimexpr#1+2.0em}}

\makeatletter
\newcount\ALG@printindent@tempcnta
\def\ALG@printindent{%
    \ifnum \theALG@nested>0
    \ifx\ALG@text\ALG@x@notext
    \else
    \unskip
    \ALG@printindent@tempcnta=1
    \loop
    \algrule[\csname ALG@ind@\the\ALG@printindent@tempcnta\endcsname]%
    \advance \ALG@printindent@tempcnta 1
    \ifnum \ALG@printindent@tempcnta<\numexpr\theALG@nested+1\relax
    \repeat
    \fi
    \fi
}%

\patchcmd{\ALG@doentity}{\noindent\hskip\ALG@tlm}{\ALG@printindent}{}{\errmessage{failed to patch}}

\AtBeginEnvironment{algorithmic}{\lineskip0pt}

\begin{document}

\date{\today}


\title{A new approach of the partial control method in chaotic systems}

\author{Rub\'en Cape\'ans}
\affiliation{Nonlinear Dynamics, Chaos and Complex Systems Group, Departamento de  F\'isica, Universidad Rey Juan Carlos, M\'ostoles, Madrid, Tulip\'an s/n, 28933, Spain}
\author{Juan Sabuco}
\affiliation{Institute for New Economic Thinking at the Oxford Martin School, Mathematical Institute, University of Oxford, Walton Well Road, Eagle House OX2 6ED, Oxford, UK.}
\author{Miguel A. F. Sanju\'{a}n}
\affiliation{Nonlinear Dynamics, Chaos and Complex Systems Group, Departamento de  F\'isica, Universidad Rey Juan Carlos, M\'ostoles, Madrid, Tulip\'an s/n, 28933, Spain}
\affiliation{Department of Applied Informatics, Kaunas University of Technology, Studentu 50-415, Kaunas LT-51368, Lithuania}

\begin{abstract}

We present here a new approach of the partial control method, which is a useful control technique applied to transient chaotic dynamics affected by a bounded noise. Usually we want to avoid the escape of these chaotic transients outside a certain region $Q$ of the phase space. For that purpose, there exists a control bound such that for controls smaller than this bound trajectories are kept in a special subset of $Q$ called the safe set. The aim of this new approach is to go further, and to compute for every point of $Q$ the minimal control bound that would keep it in $Q$. This defines a special function that we call the safety function, which can provide the necessary information to compute the safe set once we choose a particular value of the control bound. This offers a generalized method where previous known cases are included, and its use encompasses more diverse scenarios.

\end{abstract}

\maketitle

\section{Introduction}

Transient chaos is a behaviour found in nonlinear systems where trajectories behave chaotically in a certain region $Q$ of the phase space, before eventually escaping to an external attractor. In some occasions, this escape involves a highly undesirable state and therefore the application of some control scheme is required to prevent it.

Different control methods have been proposed in the literature \cite{Schwartz,Dhamala,Bertsekas,Bertsekasdos} to achieve this goal. However, these methods sometimes fail dramatically in presence of noise due to the exponential growth of small perturbations in chaotic dynamics. To deal with real systems where the presence of noise can be unavoidable, it has been recently proposed the partial control method \cite{Asymptotic,Automatic}. This method is applied on chaotic maps and it is based on the following scheme:
\begin{equation}
    \begin{array}{l}
   q_{n+1}=f(q_n)+\xi_n+u_n\\
   |\xi_n|\leq\xi_0 \hspace{0.5cm}\text{and} \hspace{0.5cm} |u_n|\leq u_0, \hspace{0.5cm} \text{with} \hspace{0.5cm} \xi_0 > u_0 > 0.
    \end{array}
\end{equation}
Here, the term  $f(q_n)$ represents the action of the map, while the terms $\xi_n$ and $u_n$ represent the disturbance and control acting on the $nth$ iteration of the map. Both, the disturbance and control are bounded so that $|\xi_n|\leq\xi_0$ and $|u_n|\leq u_0$, with $\xi_0 > u_0 > 0$. These constraints are a consequence of the limitations of both the disturbance ands control in most applications.

One of the remarkable findings of the partial control method is that, controlled trajectories exist for values $u_0 < \xi_0$. This means that the changes in the dynamics of the system induced by the disturbances can be counteracted with the application of a smaller amount of control. Such a counterintuitive result was proven in several paradigmatic systems like the H\'{e}non map~\cite{Automatic}, the Duffing oscillator~\cite{Automatic} or the Lorenz system \cite{Lorenz} as well as other models in the context of ecology or cancer dynamics~\cite{Ecology,Cancer}.

To implement this method, it is necessary to know the map $f(q)$ and the disturbance bound $\xi_0$. Then specify the region $Q$ where we want to keep the trajectories, and set the control bound $u_0$ that we want to apply. By using an algorithm called the Sculpting Algorithm \cite{Automatic} it is possible to found  the subset of points $q\in Q$ that can be controlled under the scheme (1). This subset is called the \textbf{safe set} and its shape depends on the choice of the bound $u_0$.  

When we compute the safe set, there is a minimum $u_0<\xi_0$ for which the safe set exists. That is, the safe set can be computed only for values of $u_0$ larger than this minimun $u_0$.

\begin{figure}
\includegraphics [trim=0cm 0cm 0cm 0cm, clip=true,width=0.9\textwidth]{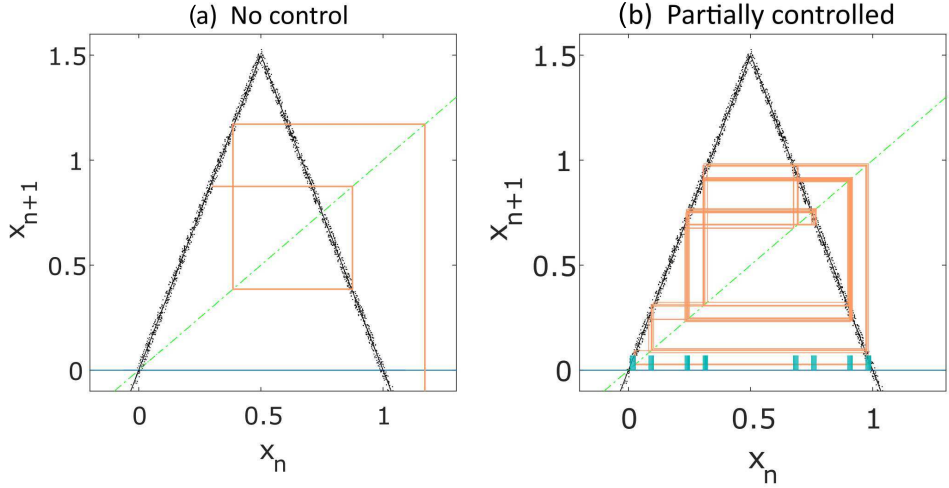}\\
\centering
\caption{\textbf{Partial control method.} In this figure the slope-3 tent map is represented. The map is affected by a uniform disturbance bounded by  $\xi_0=0.04$. The small dots help to visualize the magnitude and distribution of the disturbance. (a) An uncontrolled trajectory that escapes from the interval [0,1] after a few iterations is shown. (b) The partial control method was applied with the control bound $u_0=0.025$.  The safe set used is displayed at the bottom in green. The controlled trajectory remains in the interval $[0,1]$ when a control $|u_n|\leq 0.025$ is applied at every iteration to the $x$ variable.}
\label{1}
\end{figure}

 An example of this technique  is shown in Fig.~\ref{1}, where an uncontrolled trajectory and a controlled one are compared in the case of the slope-3 tent map. This map is given by:
\begin{equation}
     \label{}
     x_{n+1} = \left\{
	       \begin{array}{ll}
		 3 x_n + \xi_n +u_n      & \mathrm{\;for\ } x_n \le \frac{1}{2} \\
		 3(1-x_n) + \xi_n +u_n    & \mathrm{\;for\ } x_n > \frac{1}{2} \\
	       \end{array}
	     \right.
\end{equation}
where as an example, we consider a disturbance bound  $\xi_0=0.040$. In Fig.~\ref{1}(a) the control is not applied, and trajectories abandon the region $Q=[0,1]$ after a few iterations. In Fig.~\ref{1}(b) the partial control method is applied. For a $u_0=0.025$ we found the safe set displayed at the bottom of the figure. By forcing the trajectory to pass through this set, the orbit is kept in the interval $[0,1]$ by using  at every iteration a control $|u_n|\leq 0.025$.

\begin{figure}
	\includegraphics [trim=0cm -0.5cm 0cm 0cm, clip=true,width=0.6\textwidth]{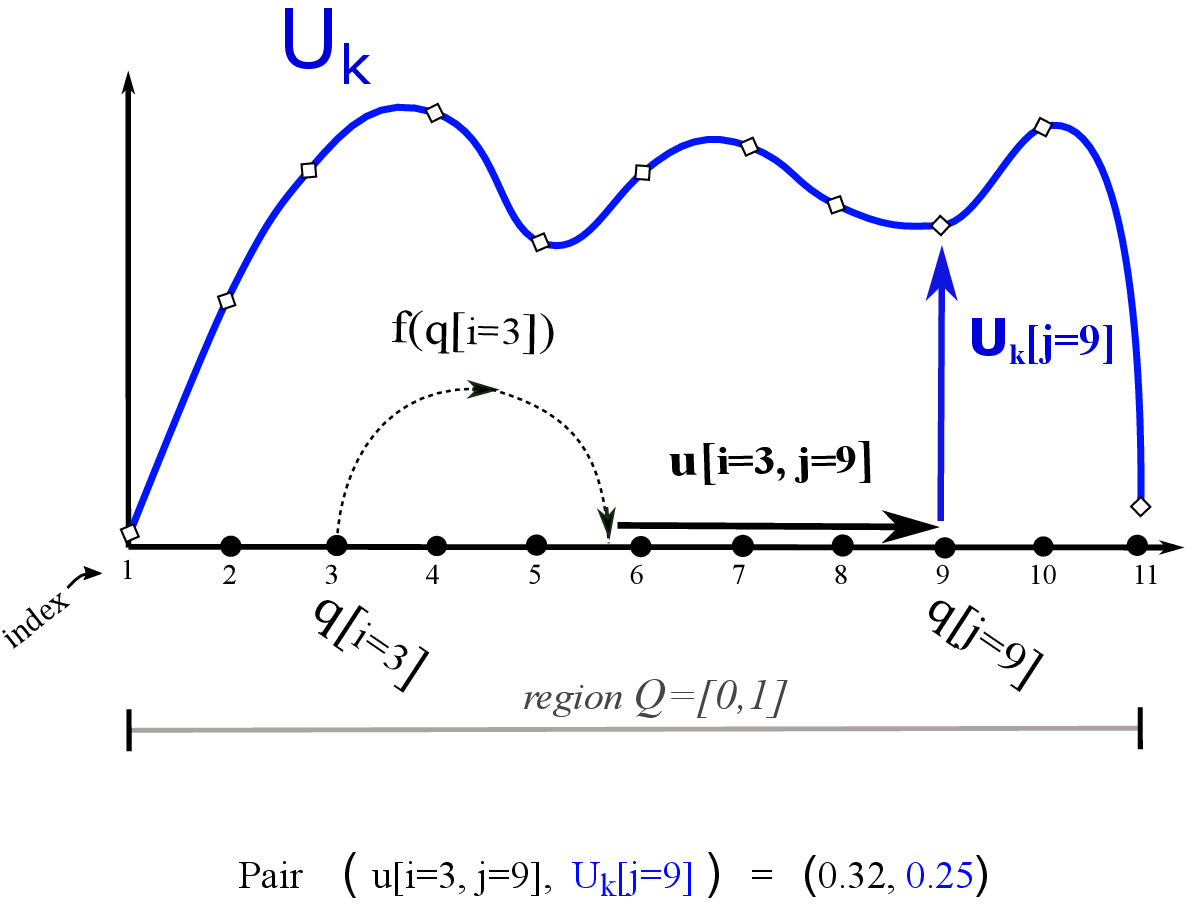}\\
	\centering
	\caption{\textbf{Definition of the  function $U_k$.} In this figure a region $Q$ (where we want to keep the dynamics) and a possible function $U_k$ (in blue) are represented. We assumed that the dynamics in this region has escapes and the control is applied to avoid these escapes. To implement the control technique,  a grid covering the region $Q$ (in this case N=11 points) was taken. The controlled dynamics is given by $q_{n+1}=f(q_n) + u_n$. We identify the starting and arrival point as $q[i]=q_n$ and $q[j]=q_{n+1}$ respectively. The control corresponding to a point $q[i]$ to go to the point $q[j]$, is denoted as $u[i,j]$, while the value $U_k[j]$ represents the control bound for the point $q[j]$ to remain in $Q$ the next $k$ iterations. Therefore the pair  of values $\big(u[i,j],\, U_k[j]\big)$ can be read as \textit{(present control, future control)}. Each possible pair, represents a choice of control. This approach evaluates all choices of control and takes the one with the minimum control bound. As an example, we have illustrated the starting point $q[i=3]$ and the choice of control $\big(u[i=3,j=9], \,U_k[j=9]\big)$  which reaches the arriving point $q[j=9]$.}
	\label{2b}
\end{figure}

\section{A new approach}

With the aim to extend the applications of the  partial control method,  a new approach has been developed. Now, the maps considered are more general and have the following form:
\begin{equation}
    \begin{array}{l}
    q_{n+1}=f(q_n, \xi_n) + u_n,
    \end{array}
\end{equation}
where $\xi_n$ is a disturbance term (random perturbation), belonging to a bounded distribution. However here, the  bound of the disturbance distribution is allowed  to be space-dependent, and can act over the variables or parameters of the map. The $u_n$ term is the control applied to the variables of the map with the aim of keeping the trajectory in the desirable region $Q$. 

To explain the goal of this approach, suppose that we start with the initial condition $q \in Q$, and in order to sustain the trajectory in $Q$ during the next $k$ iterations, we apply a sequence of control magnitudes $(|u_1|, |u_2|,.. ,|u_k|)$. However this choice of controls is not unique, and certain strategy should be followed in order to keep these controls low. What we pursue here is to find a control strategy that minimizes the upper bound (the maximum) of these controls.

To do that we define in the region $Q$ a special function that we name $U_k$.  The value $U_k(q)$ of this function represents the minimum control bound needed to sustain a trajectory (starting in $q$)  in the region $Q$ during $k$ iterations.  This means that the sequence of $k$ controls applied to this trajectory satisfy the condition $\max(|u_1|, |u_2|,.. ,|u_k|)\leq U_k(q)$. This bound is minimal, so no other controlled trajectory exists with a smaller control bound.

The $U_k, U_{k-1},..U_0$ functions implicitly define the control strategy, so we will focus here on finding these functions. This finding  is not trivial due to the chaotic dynamics present in the region $Q$.  However, it is possible to obtain them, following an  iterative procedure so that with the initial knowledge of $U_0$ we can obtain $U_1$, $U_2$.. etc.  To explain the procedure, we consider first the particular case where no disturbances are present in the controlled dynamics, and then we extend the reasoning to the case where the disturbances appear.

\begin{figure}
\includegraphics [trim=0cm -0.5cm 0cm 0cm, clip=true,width=0.9\textwidth]{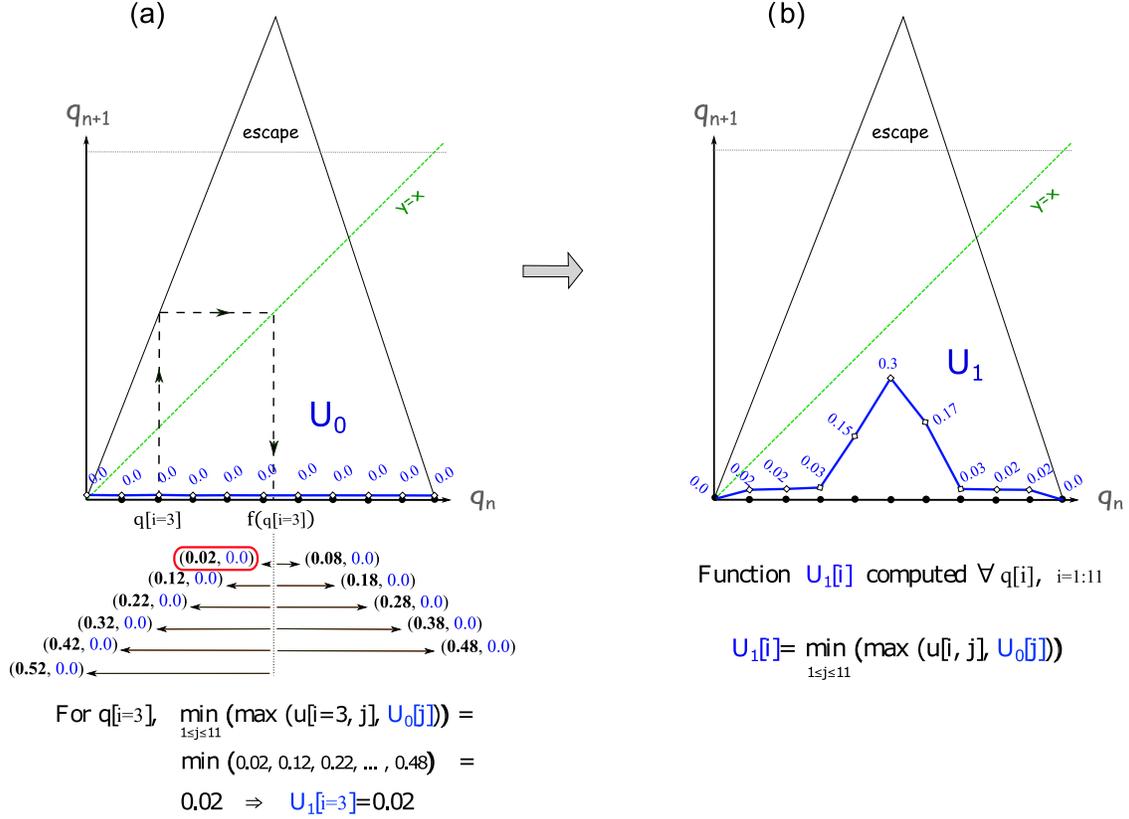}\\
\centering
\caption{\textbf{Computing the function $U_0\rightarrow U_1$ in the tent map.} (a) In the figure a tent-like map with no disturbance is represented. We have selected the interval $[0,1]$ in the region $Q$. The initial function $U_0[i]=0$, $\forall i$, is indicated in blue. Every new value $U_1[i]$ can be computed individually, and the procedure to compute $U_1[i=3]$ is shown in the figure. For that, we need to know the image $f(\,q[i=3]\,)$ and later, we compute all possible controls $u[i=3,j]$, which are represented with horizontal arrows. Afterwards, we build the corresponding pairs $(\,u[i=3,j],\, U_0[j]\,)$ indicated near the arrows respectively. The upper bound (or maximum) of each pair is indicated in bold. As we want to minimize the new control bound $U_1[i=3]$, the pair with a minimum upper bound has to be selected, that is, $U_{1}[i=3]=\min\limits_{1\leq j\leq N}\big(\max\,(\,u[i=3,j],\, U_0[j]\,)\,\big)=0.02$. (b)  The resulting function $U_1[i]$, $\forall i$ is drawn (the values indicated are approximate).}
\label{3b}
\end{figure}

\begin{figure}
\includegraphics [trim=0cm -0.5cm 0cm 0cm, clip=true,width=0.93\textwidth]{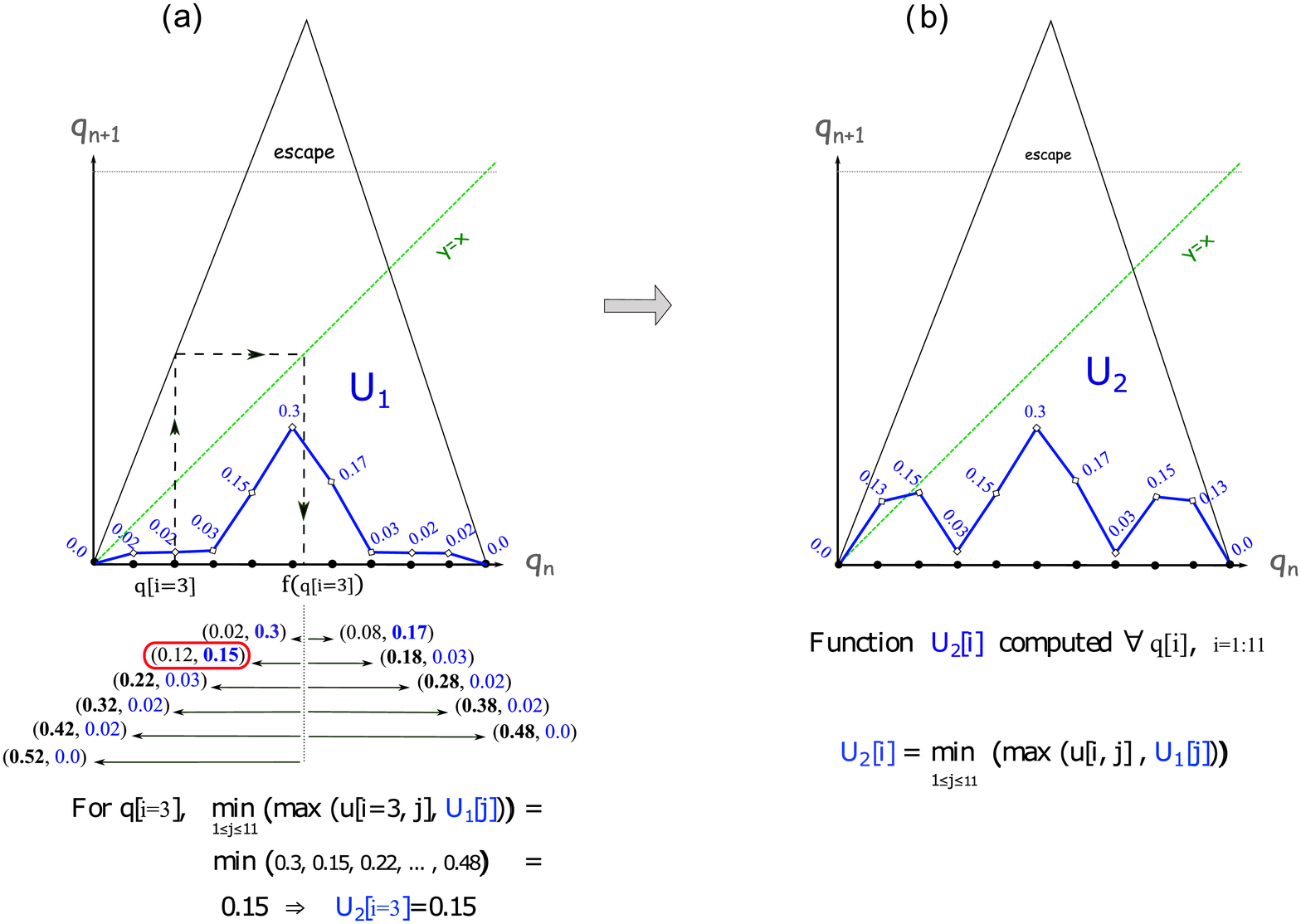}\\
\centering
\caption{\textbf{Computing the function $U_1\rightarrow U_2$ in the tent map.} (a)The previous function $U_1[i]$ is indicated in blue. Every new value $U_2[i]$ can be computed individually, and we show the procedure to compute the value $U_2[i=3]$. For that, it is necessary to know the image $f(\,q[i]\,)$ and then, compute all possible controls $u[i=3,j]$, which are represented with horizontal arrows. Then, we build all the pairs $(\,u[i=3,j],\, U_1[j]\,)$ indicated near the arrows respectively. The upper bound (or maximum) of each pair is indicated in bold. As we want to minimize the new control bound $U_2[i=3]$, the pair with a minimum upper bound has to be selected, that is, $U_{2}[i=3]=\min\limits_{1\leq j\leq N}\big(\max\,(\,u[i=3,j],\, U_1[j]\,)\,\big)$. (b)  The resulting function $U_2[i]$, $\forall i$ is drawn (the values indicated are approximate). In a similar way,  functions $U_3, U_4...$ etc. can be obtained.}
\label{4b}
\end{figure}

\subsection{Computing the functions $U_k$ in absence of disturbances.}

When no disturbances affect the system, the controlled map has the form $q_{n+1}=f(q_n) + u_n$. We use a grid on $Q$ of $N$ points, and the index $i=1:N$, to identify the starting point  $q[i]\equiv q_n$. Alternatively, we use the index $j=1:N$  to denote the  arrival point $q[j]\equiv q_{n+1}$. The controlled map in this grid takes de form $q[j]=f\big(q[i]\big)+u[i,j]$. This is illustrated in Fig.~\ref{2b}, where we have considered the interval $[0,1]$ as the region $Q$, and we have selected a grid of $N=11$ points. We show an iteration of the map, where the point $q[i=3]$ maps (control included) to the point $q[j=9]$. The particular control used is represented as $u[i=3,j=9]$. In the same figure, we also display a hypothetical function $U_k$ and its value in the arrival point $U_{k}[j=9]$. The value $u[i,j]$ represents the current control corresponding to the point i to reach the point j, while the value $U_k[j]$ represents the control bound corresponding to the point j to remain in $Q$ for the next $k$ iterations.

To illustrate the computation of the $U_k$, the slope-3 tent map shown in Fig.~\ref{3b} will be used as an example. The region $Q$ selected is the interval $[0,1]$. Note that the central points escape after one iteration. The idea is to compute recursively the functions $U_0 \rightarrow U_1 \rightarrow U_2 \rightarrow...\rightarrow U_k$. Taking into account that $U_{0}[i]$ represents control bound needed by $q[i]$ to keep its trajectory in $Q$ during $0$ iterations, it follows that $U_{0}[i]=0$, $\forall i$. This function is displayed in blue in Fig.~\ref{3b}(a). For visual convenience, both the tent map and the $U_0$ function are represented using the same axes. In the following, we will use this joint representation when the scale $axis$  overlap.

To explain how to compute $U_{1}[i]$, we take for instance, the point $q[i=3]$ shown in Fig.~\ref{3b}(a). This point maps into $f(q[i=3])$ and then, all possible controls $u[i=3,j]$ are computed, which are shown in the figure with the horizontal arrows at the bottom. For each control, the corresponding  pair $(\,u[i=3,j],\, U_0[j]\,)$ is also indicated. This pair can be read as \textit{(present control, future control)}, so that the pair that minimizes the overall control will be the pair with the minimum bound. In this case, the pair $(u \,[ i=3, j=6 ],\, U_1 \,[ j=6 ])=(0.02, 0.0)$ marked in red has the minimum bound $U_1[i=3]=0.02$. This value represents the minimum upper control bound for just one iteration. In general, the values of the function $U_1$ can be found as $U_1[i]=\min\limits_{1\leq j\leq 11}\big(\,\max\,(\,u[i,j], U_0[j]\,)\big)$.

  The resulting function $U_1$ is displayed in Fig.~\ref{3b}(b). It can be seen that the central points of $Q$ maps outside $Q$, and therefore they need a big control to return to $Q$ in just one iteration. Therefore a central peak appears in the function $U_1$. 

\begin{figure}
	\includegraphics [trim=0cm -0.5cm 0cm 0cm, clip=true,width=0.75\textwidth]{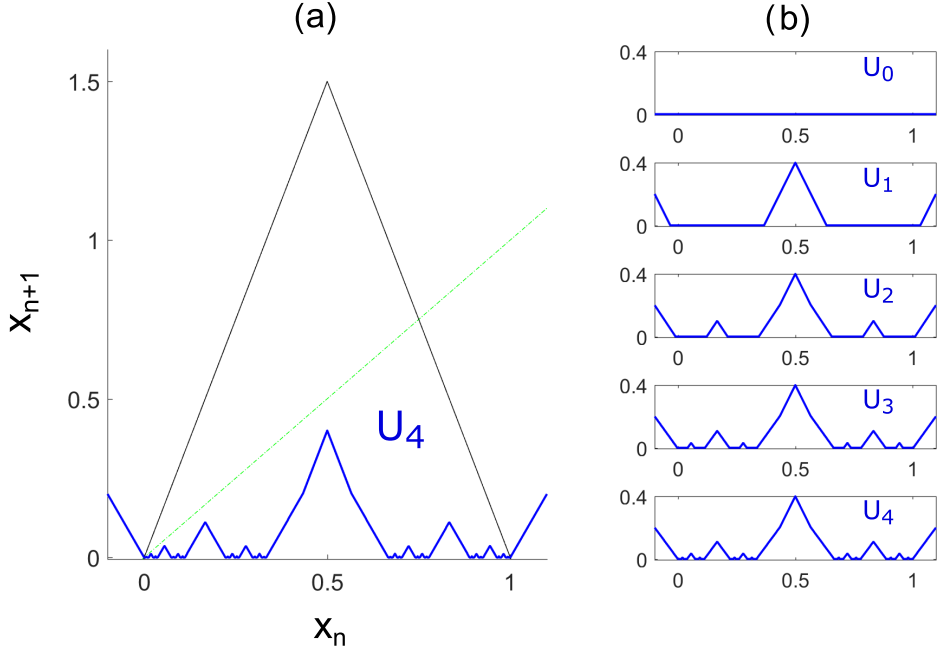}\\
	\centering
	\caption{\textbf{Functions $U_0\rightarrow U_1\rightarrow U_2 \rightarrow U_3 \rightarrow U_4$ in the slope-3 tent map.} (a) The slope-3 tent map where the region $Q$ selected is the interval $[-0.1,1.1]$. Taking an uniform grid of $1000$, the  function $U_4$ (in blue) was computed. (b) The successive functions $U_k$ (starting with $U_0$) that have been computed to obtain  $U_4$.}
	\label{5b}
\end{figure}

Once we have  $U_1$, the function $U_2$ can be computed following the same process (see Fig.~\ref{4b}). Taking again the initial point $q[i=3]$, the action of the tent map $f(\,q[i=3]\,)$ is shown in the figure. Then a control $u[i=3,j]$ is applied. All  possible  pairs $(u[i=3,j], U_1[j])$ are indicated. In this case,  the pair $(u \,[ i=3, j=5 ],\, U_1 \,[ j=5 ])=(0.12, 0.15)$ marked in red has the minimum bound (0.15). This value represents the minimum upper control bound for $2$ iterations. Therefore $U_2[i=3]=0.15$. In general the values of the function $U_2$ can be found as $U_2[i]=\min\limits_{1\leq j\leq 11}\big(\,\max\,(\,u[i,j], U_1[j]\,)\big)$. In Fig.~\ref{4b}(b) the function $U_2$ is shown.

Equivalently, we compute $U_3$, $U_4$... etc. In general, in absence of any disturbance, we have the following recursive formula to compute the  functions $U_k$:

\vspace{0.5cm}

\begin{equation}
	\begin{array}{l}
	   \tcbhighmath{U_{k+1}[i]=\min\limits_{1\leq j\leq N}\big(\max\,(\,u[i,j],\, U_k[j]\,)\,\big)}\\\\
	   i\equiv \text{index of the starting point~~} q[i],~ i=1:N.\\
	   \hspace{0.75cm} \text{where } N=\text{total number of grid points}.\\\\
	   j\equiv \text{index of the arrival point~~} q[j],~ j=1:N
	\end{array}
	\vspace{0.5cm}
\end{equation}
starting with $U_0[i]=0 \,\,\,\, \forall i$. Note that the values of $u[i,j]$ remain unchanged for every iteration of the algorithm, so they only need to be calculated once. In Fig.~\ref{5b}, we display the process for the slope-3 tent map. The region $Q$ has been selected to be the interval $[-0.1,1.1]$. We have used a uniform grid of $1000$ points in this interval. On the right side of the figure, the successive functions $U_0 \rightarrow U_1 \rightarrow U_2 \rightarrow U_3 \rightarrow U_4$ are shown.


\subsection{Computing the functions $U_k$ in presence of disturbances}

The extension of the recursive algorithm in the case of systems affected by disturbances is rather straightforward. Now the dynamics is given by $q_{n+1}=f(q_n,\xi_n) + u_n$, where $\xi_n$ is the disturbance term belonging to a bounded distribution. 

\begin{figure}
	\includegraphics [trim=0cm -0.5cm 0cm 0cm, clip=true,width=0.4\textwidth]{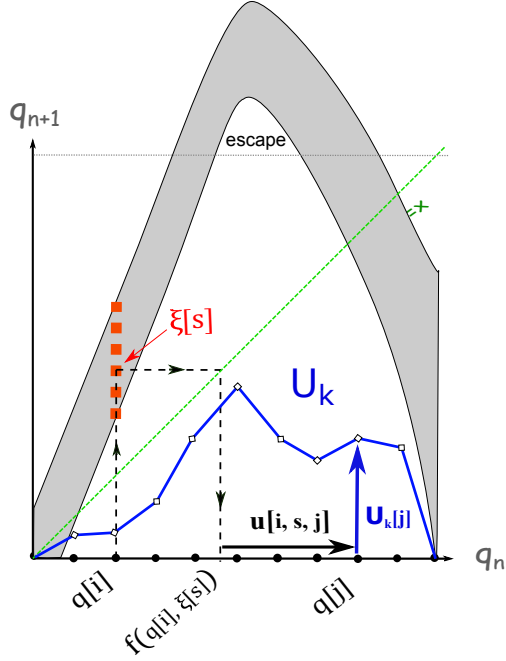}\\
	\centering
	\caption{\textbf{Scheme of a map affected by a bounded disturbance distribution.} The extension of the algorithm in the case of maps affected by a bounded disturbance distribution, is rather straightforward. In this case, given a point $q[i]$, to compute the upper control bound $U_{k+1}[i]$, we have to consider all disturbed images $f(q[i],\xi[s])$. Then compute all the corresponding control bounds as in the case of no disturbances, and finally extract the maximum control among them all.}
	\label{6b}
\end{figure}

In Fig.~\ref{6b}, we illustrate the case of a map affected by a bounded disturbance distribution. The main complication here is that, due to the disturbance, the same point has multiple disturbed images. This number can be infinite and therefore, a discretization must be taken to perform the computations (see the red dots in Fig.~\ref{6b}).  Given a point $q[i]$, we denote the grid of possible images as $f(q[i], \xi[s])$, where $s=1:M_i$ is the index of every individual disturbance. The number of disturbed images $M_i$ can take different values depending on the particular point $q[i]$. The control corresponding to the point $q[i]$ and affected by the disturbance $\xi[s]$, to reach the point $q[j]$, is denoted as $u[i,s,j]$.

\begin{figure}
	\includegraphics [trim=0cm -0.5cm 0cm 0cm, clip=true,width=0.9\textwidth]{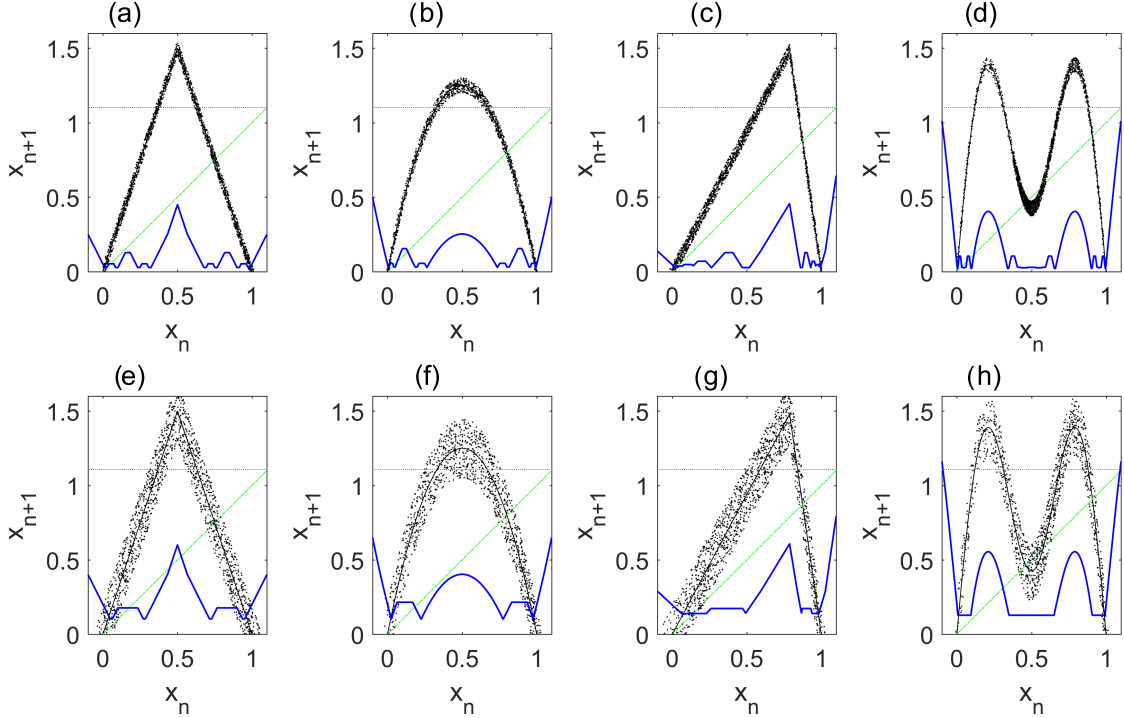}\\
	\centering
	\caption{\textbf{Safety function $U_\infty$ for different maps affected by a disturbance.} This figure shows how the safety function (in blue) changes depending on the map and  the disturbance affecting it. In all cases, the convergence of the safety functions was achieved with $15$ iterations or less of the algorithm. The maps represented are the following: (a,e) Tent map. (b,f) Logistic map. (c,g) Asymmetric tent map. (d,h) Map with two symmetric hills. The horizontal grey line at $x_{n+1}=1.1$ indicates the escape. Points $q$ that map above this line, escape directly from the region $Q=[-0.1,1.1]$. The figures on the top (a,b,c,d) are affected by a uniform disturbance distribution bounded by $\xi_0=0.05$. In contrast, for the maps at the bottom (e,f,g,h), the disturbance bound is $\xi_0=0.2$.  Note that the safety functions for the bottom maps take larger values, due to the larger disturbances affecting them.}
	\label{7b}
\end{figure}

Now, to compute the functions $U_k$ in presence of disturbances, we follow a similar reasoning as in the case where there are no disturbances. However, in this case we also have to evaluate all the disturbed images for a given point $q[i]$, and take the maximum among them all to obtain an overall upper control bound. Therefore, the recursive formula in presence of disturbances is given by:\\
\begin{equation}
\begin{array}{l}
\tcbhighmath{U_{k+1}[i]=\max\limits_{1\leq s\leq M_i}\Big(\min\limits_{1\leq j\leq N}\big(\max\, (\,u[i,s,j],\, U_k[j]\,)\,\big)\,\Big)} \\\\
i\equiv \text{index of the starting point~~} q[i],~ i=1:N.\\
\hspace{0.75cm} \text{where } N=\text{total number of grid points}.\\\\
s\equiv \text{index of the disturbance~~~} \xi[s],~ s=1:M_i.\\
\hspace{0.75cm} \text{where } M_i=\text{number of disturbed images corresponding with } q[i].\\\\
j\equiv \text{index of the arrival point~~} q[j],~ j=1:N

\end{array}
\vspace{0.5cm}
\end{equation}
starting with $U_0[i]=0 \,\,\,\,\, ,\forall i$. Note that in this iterative formula, the $u[i,s,j]$ values remain unchanged every iteration of the algorithm. Thus, they  only need to be calculated once.

\section{The safety function $U_\infty$ and the safe sets.}

In this section, we study the important case where the goal of the controller is to keep the trajectory in the region $Q$ forever with the smallest control bound. To do that, it is required to find $U_\infty$ (that we call the \textbf{\textit{safety function}}) and therefore iterate infinite times the algorithm. However, if the algorithm converges for a given iteration $k$ so that  $U_{k+1}=U_k$, then it follows that  $U_\infty =U_k$  and the iterative process is finished.  We do not intend here to explore the necessary mathematical conditions to achieve the convergence. Our finding is that for the analyzed transient chaotic maps, the algorithm converges in a few iterations. In the next sections some examples supporting this point will be provided.

To show different examples of safety functions and how the  disturbances affect them, we represent in Fig.~\ref{7b} the safety function $U_\infty$ (in blue) for different maps. The maps at the top (a,b,c,d) are affected by the same disturbance bound $\xi_0=0.05$. The maps at the bottom (e,f,g,h) are the same respectively, but affected by a bigger disturbance bound ($\xi_0=0.2$). Note that the safety function $U_\infty$ has larger values in this case, since a larger control bound is needed to sustain trajectories affected by larger disturbances. 

\begin{figure}
	\includegraphics [trim=0cm -0.5cm 0cm 0cm, clip=true,width=0.97\textwidth]{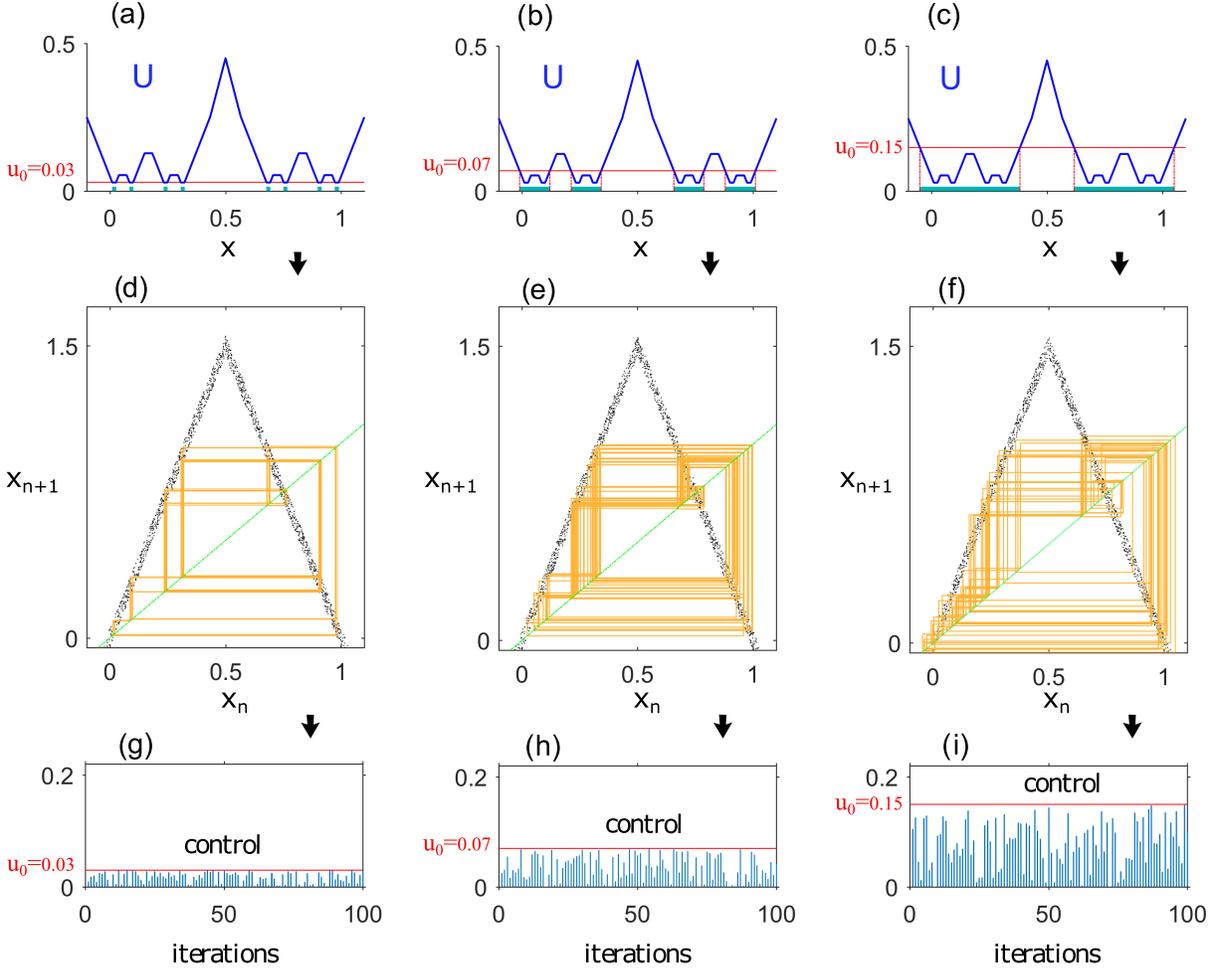}\\
	\centering
	\caption{\textbf{Extracting the safe sets from the  safety function.} Once the safety function  $U_\infty$ is computed,  the safe set is the set of points $q \in Q$ that satisfy $U_\infty(q)\leq u_0$, where $u_0$ is the control bound that we want to apply. In panel (a) we draw the safety function (in blue) corresponding to the map of Fig.~\ref{7b}(a), the tent map affected by an uniform disturbance with bound $\xi_0=0.05$. The safe set corresponding to $u_0=0.03$  is shown at the bottom (green bars). In panel (d) a trajectory is controlled by applying every iteration a control $|u_n|\leq u_0$ that forces the trajectory to pass through the safe set. The controls $|u_n|$ applied are shown in  panel (g), and for convenience we only display the first $100$ iterations. Panels (b,e,h) are equivalent but taking instead a control bound $u_0=0.07$. In panels (c,f,i) the control bound is $u_0=0.15$. Note that the larger the $u_0$ value, the larger the safe set, and therefore the trajectory is allowed to explore more points of the $Q=[-0.1,1.1]$ region.}
	\label{8b}
\end{figure}

To control a trajectory by means of the safety function, we need to specify first the upper control bound $u_0$ that we want to apply.  This value must be chosen so that $u_0\geq \min(U_\infty)$. Above this minimum, any value $u_0$ is allowed. The  set of points $q$ for which $U_\infty(q)\leq u_0$ constitutes what we call the \textit{safe set}. Only this set of points can be controlled forever by applying  controls $|u_n|\leq u_0$, where in each iteration of the map, $u_n$ is chosen to force the trajectory to pass through the safe set.  Very often, the choice of the control $u_n$ is not unique and therefore multiple controlled trajectories are possible. This makes the method very flexible

In Figs.~\ref{8b}(a-b-c) the safety functions corresponding to the maps of the Fig.~\ref{7b}(a) are shown. Different control bounds $u_0$ were taken and at the bottom the respective safe sets have been drawn. For each value $u_0$, a particular controlled trajectory is shown in Figs.~\ref{8b}(d-e-f). For clarity, we display only  $100$ iterations of the trajectory. The control applied in every iteration of these trajectories is represented at the bottom of Figs.~\ref{8b}(g-h-i) respectively.

\subsection{Application to the tent map affected by asymmetric disturbances}

\begin{figure}
	\includegraphics [trim=0cm 0cm 0cm 5cm, clip=true,width=0.98\textwidth]{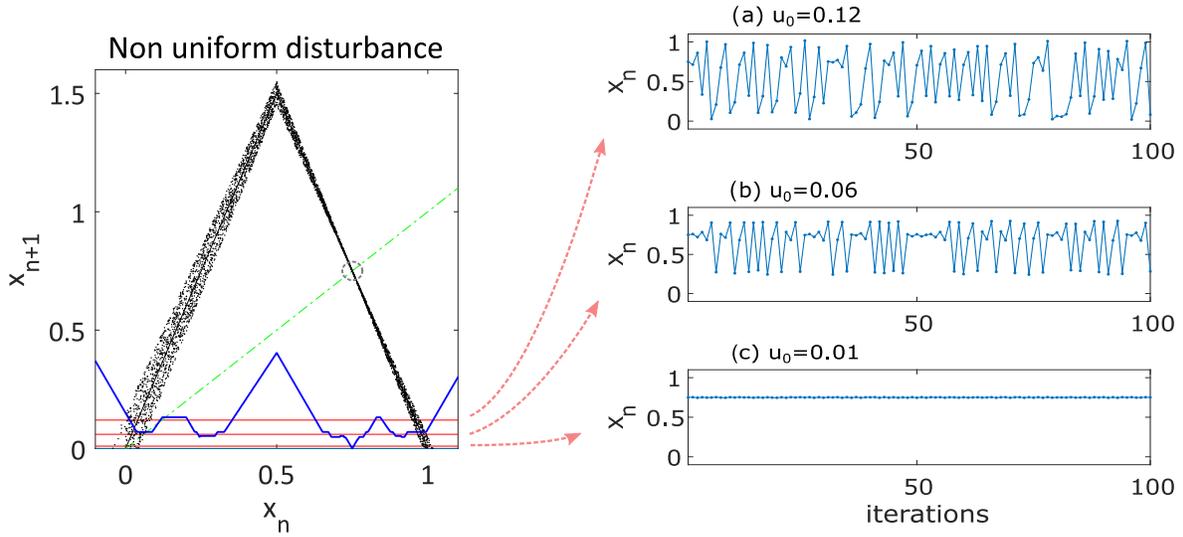}\\
	\centering
	\caption{\textbf{Asymmetric disturbance affecting the map.}  On the left panel we show the slope-3 tent map affected by an asymmetric disturbance. In particular the disturbance was set to have a zero value in the fixed point $x^*=0.75$ highlighted with a circle. This choice was made on purpose to show that the safety function $U_\infty$ (in blue) has a minimum in this point, since no control is needed to keep the trajectory in the fixed point due to the zero disturbance affecting it. Three different control bounds $u_0$ (red horizontal bounds) have been tested. On the right panel we represent the corresponding controlled trajectories.  Depending on the control bound the qualitative behavior changes drastically. A small control keeps the trajectory around the fixed point, while larger $u_0$ values let the trajectory explore other points of the region [-0.1,1.1].}
	\label{5}
\end{figure}

In the previous examples, we have considered maps where the disturbance $\xi_n$ affecting the trajectories were uniformly bounded so that $|\xi_n|\leq \xi_0$ $\forall x \in Q$.
However there is no impediment to apply the algorithm in case of non-uniform disturbance bounds. To show an example, we consider the slope-3 tent map affected by a non-uniform disturbance. The system is given by:
\begin{equation}
\label{}
x_{n+1} = \left\{
\begin{array}{ll}
3 x_n + \xi_n(4x_n-3) +u_n      & \mathrm{\;for\ } x_n \le \frac{1}{2} \\
3(1-x_n) + \xi_n(4x_n-3) +u_n    & \mathrm{\;for\ } x_n > \frac{1}{2}, \\
\end{array}
\right.
\end{equation}
where the term $\xi_n(4x_n-3)$ models the asymmetric disturbance distribution (see Fig.~\ref{5}). This particular choice of disturbance was made on purpose to show the particular shape of the function $U_\infty$. For this map, the fixed point  $x^*=0.75$ is affected by a zero disturbance, and therefore it needs zero control since $f(x^*)=x^*$. For this reason, we expect that the safety function evaluated in the fixed point takes the value $U_\infty(x^*)=0$. 

We have chosen a uniform grid of 1000 points in the region $Q=[-0.1,1.1]$, and we have computed the safety function $U_\infty$, which is shown in Fig.~\ref{5}. We can observe that $U_\infty$  has a minimum in the fixed point $x=0.75$. This minimum control is virtually zero, as we expected. In the right panel of Fig.~\ref{5} different controlled trajectories are displayed for increasing control values $u_0$. Note that with the control bounds $u_0=0.12$ and $u_0=0.06$ the trajectory behaves chaotically (affected by the disturbances), while in the case of $u_0=0.01$, the trajectory remains in the fixed point. This interesting result could be used by the controller to change the qualitative behavior of the trajectory, just varying the control value $u_0$.

\subsection{Application to the H\'enon map}

\begin{figure}
	\includegraphics [trim=0cm 0cm 0cm 0cm, clip=true,width=0.5\textwidth]{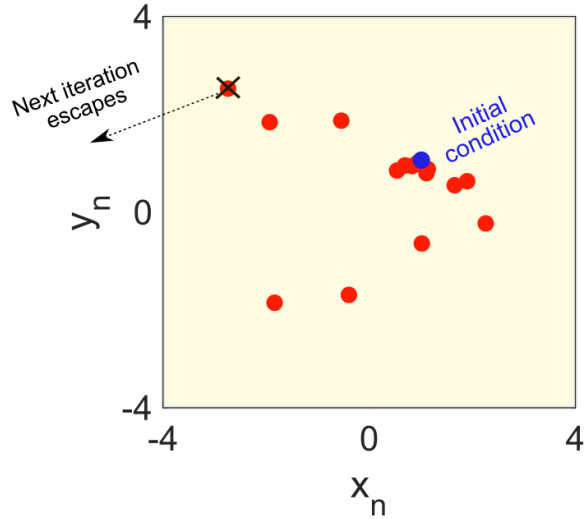}\\
	\centering
	\caption{\textbf{Uncontrolled trajectory in the H\'{e}non map.} The H\'{e}non map for the parameter values $a=2.16$ and $b=0.3$ and affected by a uniform disturbance bounded with $\xi_0=0.1$. The blue dot is the initial condition and thr reds dots describe a chaotic transient path.an eventually escapes. The dot marked with a cross is the last iteration of the trajectory in the square $Q=[-4,4] \times [-4,4]$, and the next one escapes from $Q$.}
	\label{6}
\end{figure}

\begin{figure}
	\includegraphics [trim=0cm 0cm 0cm 0cm, clip=true,width=0.62\textwidth]{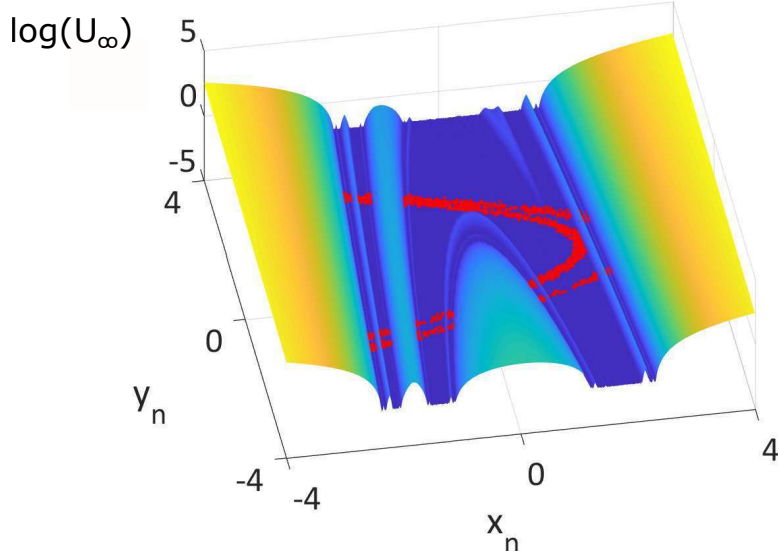}\\
	\centering
	\caption{\textbf{The 2D safety function for the H\'{e}non map.} Taking a uniform disturbance bounded by $\xi_0=0.1$ and with the goal of keeping the trajectory in the square $Q=[-4,4] \times [-4,4]$, the safety function $U_\infty$ has been computed. The algorithm takes 13 iterations to converges in a  grid of $2000\times2000$ points. This function has a minimum value of $0.08$. The logarithm of $U_\infty$ is shown here to enhance the visualization. We represent a controlled trajectory (in red) with a control bound $u_0=0.08$.  This trajectory never abandons the square $Q=[-4,4] \times [-4,4]$.}
	\label{7}
\end{figure}

In order to compute a two-dimensional safety function, we use here the H\'enon map, defined as:
\begin{equation}
\begin{array}{l}
x_{n+1}=a-b y_n- x_n^2  \\
y_{n+1}=x_n.   \\
\end{array}
\end{equation}
This map shows transient chaos for a wide range of parameters $a$ and $b$. Here we have chosen the parameter values $a=2.16$ and $b=0.3$. For these values, the trajectories with initial conditions in the square $[-4,4] \times [-4,4]$ have a short chaotic transient, before finally escaping this region towards infinity (see Fig.~\ref{6}).

 In this example, we consider a situation where the variables $(x,y)$ are affected by a uniform and bounded disturbance $(\xi^x_n, \xi^y_n)$  so that $\parallel\xi^x_n, \xi^y_n\parallel= \sqrt{(\xi^x_n)^2+(\xi^y_n)^2}\leq \xi_0$. To keep the orbits in $Q=[-4,4] \times [-4,4]$, we apply a control $(u^x_n,u^y_n)$ also bounded $\parallel u^x_n, u^y_n\parallel \leq u_0$. The controlled dynamics of the system is then given by:
\begin{equation}
\begin{array}{ l }
x_{n+1}=a-b y_n-x_n^2+\xi^x_n+u^x_n  \\
y_{n+1}=x_n +\xi^y_n+u^y_n. \\
\end{array}
\end{equation}

We have applied the extended partial control algorithm with a disturbance bound $\xi_0=0.10$, obtaining the safety function $U_\infty$ shown in Fig.~\ref{7}. The logarithm of $U_\infty$  has been plotted for a better visualization. The minimum of $U_\infty$ is found at the value 0.07. In the figure it has been represented a controlled trajectory (red dots) obtained by setting a control bound $u_0=0.08$.   The controlled trajectory  remains in the square $[-4,4]$ forever.

\subsection{Application to a time series from an ecological system.}

In this example, we have worked with an ecological model that describes the interaction between 3 species: resources, consumers and predators. The interest of this model lies in the fact that, for some choices of parameters, transient chaos appears involving the extinction of one of the species.  Without no control, the system evolves from a situation where the three species coexist towards a state where just two species survive, while predators get extinct. 

\begin{figure}
	\includegraphics [trim=0cm 0cm 0cm 0cm, clip=true,width=0.7\textwidth]{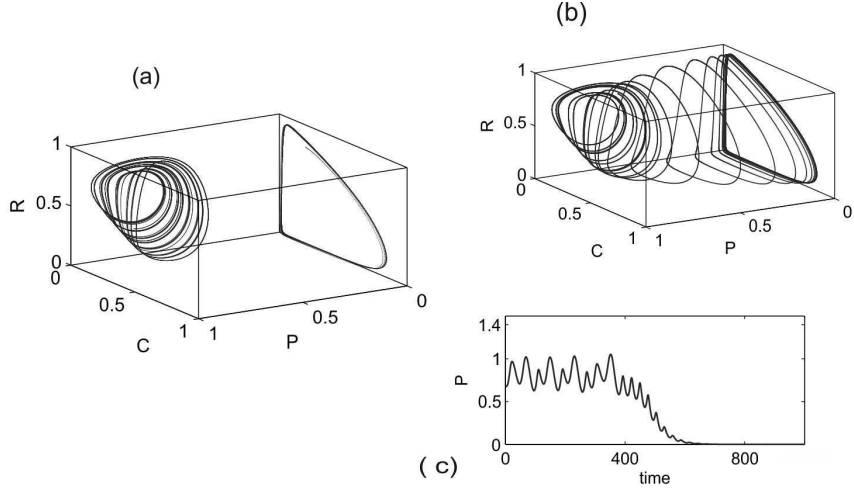}\\
	\centering
	\caption{\textbf{Dynamics of the extended McCann-Yodzis (Eqs.~\ref{eqY})}. Depending on the values of the parameters  different dynamics are possible. (a) Before the boundary crisis ($K=0.99$, $\sigma=0$), there are two possible attractors depending on the initial conditions: one chaotic attractor where the three species coexist, and one limit cycle where only the resources and consumers coexist. (b) The case treated here, for values ($K=0.99$, $\sigma=0.07$), a chaotic crisis appears and the limit cycle is the only asymptotic attractor. (c) Time series of the predators population corresponding to the case $(b)$. The predators eventually get extinct.}
	\label{8}
\end{figure}

The model that we have used is an extension of the McCann-Yodzis model \cite{McCann} proposed by Duarte et al. \cite{McCann,Duarte}, which describes the dynamics of the population density of a resource species $R$, a consumer $C$ and a predator $P$. The resulting model is given by the following set of nonlinear differential equations:
\begin{eqnarray}
\frac{dR}{dt}&=&R\left(1-\frac{R}{K}\right )-\frac{x_c y_c CR}{R+R_0}  \nonumber \\
\frac{dC}{dt}&=&x_c C\left(\frac{y_c R}{R+R_0}-1\right)-\psi(P) \frac{y_p C}{C+C_0} \label{eqY} \\
\frac{dP}{dt}&=&\psi(P) \frac{y_p C}{C+C_0}-x_p P. \nonumber
\end{eqnarray}
Depending on the parameters values, different dynamical behaviours can be found (see Fig.~\ref{8}). Following \cite{Duarte} we have fixed the model parameters : $x_c=0.4$, $y_c=2.009$, $x_p=0.08 $, $y_p=2.876 $, $R_0=0.16129$, $C_0=0.5$, $K=0.99$ and $\sigma=0.07$. For these values transient chaos appears, and the predators eventually get extinct as  shown in  Figs.~\ref{8}(b) and ~\ref{8}(c).

\begin{figure}
	\includegraphics [trim=0cm 0cm 0cm 0cm, clip=true,width=0.9\textwidth]{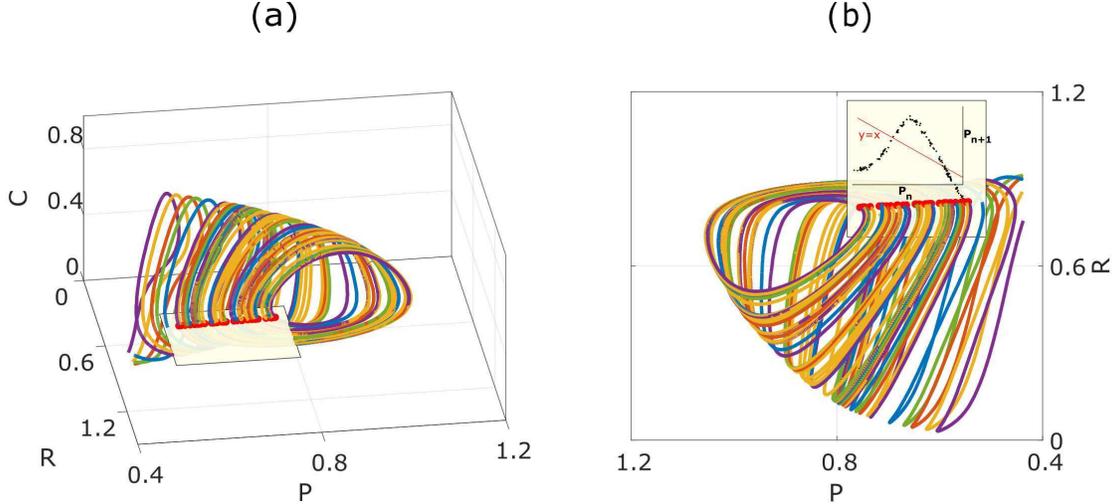}\\
	\centering
	\caption{\textbf{Building the map from several trajectories.} It is possible to discretize the dynamics of the ecological model by taking a Poincar\'{e} section. In this case, we have chosen the section with $C=0.24$ as  shown on the left panel. With the set of points $(R_n, C_n, P_n)$ intersecting the plane, it is possible to build a return map of the form $P_{n+1}=f(P_n)$ as  represented on the right panel. As the trayectories escape towards values $P \rightarrow 0$ after a short transient, several trajectories (represented with different colors) were taken to build an accurate return map. For this choice of the Poincaré section the values $R_n$ and $C_n$ in the Poincar\'{e} section remain practically constant so that only the values $P_n$ will be controlled.}
	\label{9}
\end{figure}

With the aim of avoiding the extinction, we have computed the safety function. To do that, first we have discretized the dynamics to obtain a map. It is straightforward to build a map taking a Poincar\'{e} section that intersects the flow. In this case, we have chosen the plane $C=0.24$ as shown in Fig.~\ref{9}(a). For this Poincar\'{e} section the intersection of the plane and the flow, gives us a set of points $(R_n, C_n, P_n)$ that is approximately one-dimensional. Note that $C_n$ has a constant value equal to 0.24, and the variable  $R_n$ is practically constant. Therefore it is possible to construct a return map of the form $(P_n, P_{n+1})$ and control the system just perturbing the variable $P_n$. Due to the finite escape time of the transient chaotic trajectories, several trajectories were simulated (displayed with different colors in Fig.~\ref{9}) to obtain a representative return map.

We consider here two different cases. First, a situation where the trajectories are affected by continuous noise in the variables. Second, the case where a continuous noise is affecting the parameter $K$ of the system. We want to point out here the difference between the meanings of disturbance and noise.  In our convention, the disturbance term only appears in the map and represents the amount of uncertainty measured in this map. In this sense, the disturbance is the product of the accumulated noise along the trajectory during one iteration of the map. The controlled scheme is given by:

\begin{eqnarray}
P_{n+1}=f(P_n,\xi_n)+u_n,
\end{eqnarray}
where $\xi_n$ is a particular disturbance whose bound $\xi_0$ may be space-dependent.

\begin{figure}
	\includegraphics [trim=0cm 0cm 0cm 0cm, clip=true,width=0.9\textwidth]{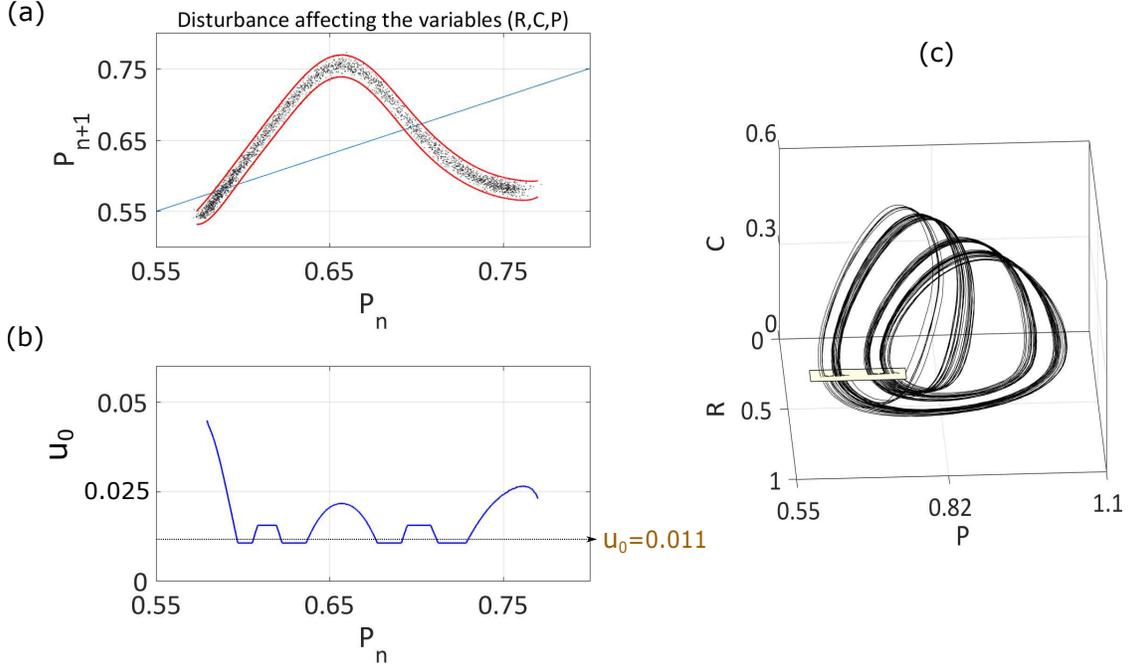}\\
	\centering
	\caption{\textbf{Continuous noise affecting the variables.} (a) Return map obtained by means of 3000 intersections of the trajectories with the Poincar\'{e} section. A continuous noise is affecting the variables $(R,C,P)$ and it arises in the return map as a stripe. Red lines represent the quantile regression calculated for quantiles 0.01 and 0.99. The gap between the red lines represent the disturbance bound. In this case the gap is rather uniform in all the map. (b) Taking the region $Q$ as the interval $[0.58,0.76]$, the safety function $U_\infty$ (in blue) has been computed obtaining a minimum value of $0.010$. (c) A controlled trajectory has been computed with a control bound of $u_0=0.011$. Every time the trajectory crosses the section, a control $|u_n|\leq 0.011$ is applied to put the orbit again in the nearest point $P_n$ with $U_\infty(P_n) \leq 0.011$.}
	\label{10}
\end{figure}

In the first scenario, the trajectories were obtained by using a RK4 integrator with a Gaussian noise affecting the variables $(R,C,P)$. In Fig.~\ref{10}(a) the return map obtained via 3000 intersections of the trajectories with the Poincar\'{e} section is shown. With these points it is possible to reconstruct the map including the disturbance. Note that in this sense, noise removal techniques are useless here since we want to include the disturbances (the accumulated noise measure in the map). To do that, different statistical techniques can be used. One very powerful is the bootstrapping technique that allows the  estimation of the sampling distribution of almost any statistic using random sampling methods. However for simplicity, we use here a quantile regression technique to estimate the upper and lower bounds of the map. Taking the quantile values 0.01 (lower bound) and 0.99 (upper bound) we obtain the two red curves shown in Fig.~\ref{10}(a).  The gap between the two curves contains the disturbed points corresponding to each $P_n$ value. We can see that the disturbance gap is rather uniform in this case.

In order to avoid the extinction of predators, the  region $Q$ selected to keep the trajectory is the interval $[0.58,0.76]$, where a grid of $2000$ points were taken for the computations. Then, we have computed the safety function $U_\infty$  shown  in Fig.~\ref{10}(b). The minimum of this function corresponds to the value 0.010. Taking a control bound $u_0=0.011$ a trajectory was controlled using the corresponding safe set. Only the variable $P_n$ needs to be controlled since $R_n$ and $C_n$ remain practically constant. In Fig.~\ref{10}(c), 500 iterations of the controlled trajectory are displayed. Every time the Poincar\'{e} section is crossed, a suitable control $|u_n|\leq 0.011$ is applied. As a result, the extinction of the predators is avoided and the 3 species coexist in a stable chaotic regime.

\begin{figure}
	\includegraphics [trim=0cm 0cm 0cm 0cm, clip=true,width=0.9\textwidth]{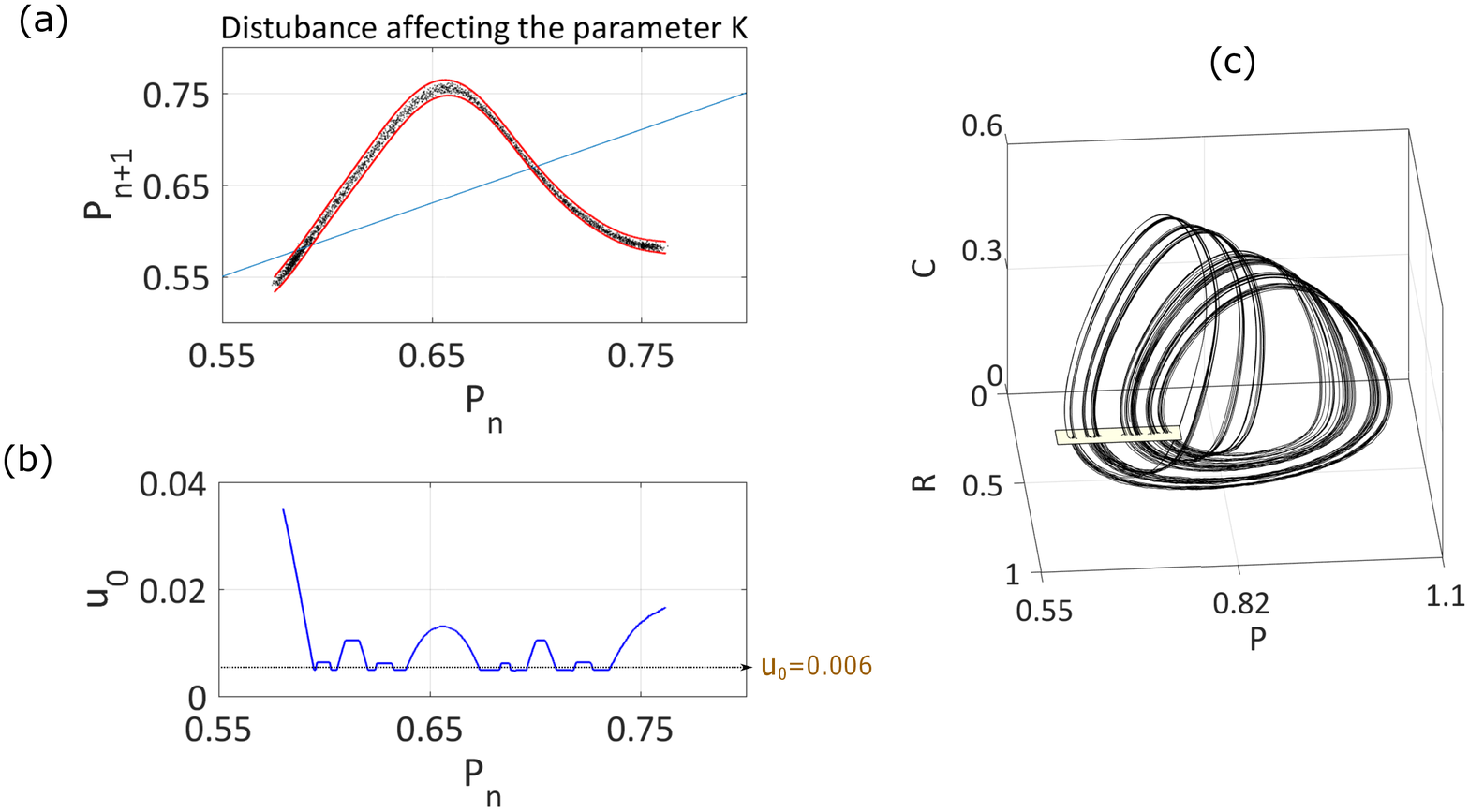}\\
	\centering
	\caption{\textbf{Continuous noise affecting the parameter $K$.} (a) Return map obtained by means of 3000 intersections of the trajectories with the Poincar\'{e} section. A continuous noise is affecting the parameter $K$ and it arises as a stripe in the return map. Red lines represent the quantile regression calculated for quantiles 0.01 and 0.99. The gap between the red lines represent the bound of the disturbance. The gap in this case is not uniform, since some points $P_n$ are affected by bigger disturbances than others. (b) Taking the region $Q$ as the interval $[0.58,0.76]$, the safety function $U_\infty$ (in blue) has been computed obtaining a minimum value of $0.005$. (c) A controlled trajectory was computed with a control bound of $u_0=0.006$. Every time the trajectory crosses the section, a control $|u_n|\leq 0.006$ is applied to put the orbit again in the nearest point $P_n$, with $U_\infty(P_n) \leq 0.006$.}
	\label{11}
\end{figure}

In the second situation, we consider a small Gaussian noise affecting the parameter $K$ of the system. This noise affects continuously $K$ and it has been included in the integrator. Proceeding in a similar way to the previous case, we obtain the return map shown in Fig.~\ref{11}(a). It can be appreciated that, in comparison with the first scenario, the disturbance interval (gap between red lines) is smaller and less uniform. Therefore the $U_\infty$ function, which is shown in Fig.~\ref{11}(b) is quite different.  In the Fig.~\ref{11}(c) a controlled trajectory is displayed, for which we have used a control bound $u_0=0.006$. We have only used 500 iterations to represent the controlled trajectory. During these iterations, no control $u_n$ exceeds the control bound $u_0$. However, due to the Gaussian noise (not bounded) affecting the parameter $K$, it may happen that at certain iteration we need an extra control. For example, if we work with a  map affected by a normal disturbance distribution, and we bound it with a three-sigma interval, the safety function $U_\infty$ obtained and the upper bound $u_0$ selected,  will be valid the $99.7\%$ of the times. The rest of iterations ($0,3\%$), a suitable control will minimize the risk of having to apply a big control in the following iterations. As we know \textit{how safe} is every point $q \in Q$, this suitable control can be chosen efficiently.

\section{Conclusions}

We have presented here a new algorithm in the context of the partial control method. This method is applied to maps of the form $q_{n+1}=f(q_n, \xi_n) +u_n$,  where $\xi_n$ is the disturbance and $u_n$ the control. Given a region $Q$ where the dynamics presents an escape, the method calculates directly the minimum   control bound  needed to sustain a trajectory  in the region $Q$ forever. To do that, we have introduced the safety function $U_\infty$ that can be computed through a recursive algorithm. This function characterizes every state $q\in Q$ and tell us how much effort is required to control it.  Once the safety function is computed, we only need  to pick a bound $u_0 \geq \text{min}(U_\infty)$. Controlled trajectories are possible by applying a suitable control $|u_n|\leq u_0$ every iteration.

The new partial control algorithm has been proven in the one-dimensional tent map and the two-dimensional H\'enon map, under a non-uniform and a uniform disturbance bound respectively. We have also applied the control method to a continuous ecological system where one of the species eventually gets extinct via a boundary crisis. Two different scenarios were studied, a continuous noise affecting the variables, and a continuous noise affecting one parameter of the system. In both cases the safety function $U_\infty$ was obtained and the trajectories controlled, avoiding the extinction. 

We show that the use of the safety functions $U_\infty$  makes this partial control approach very robust and specially useful in the case of experimental time series. Although the method was presented here to avoid undesirable escapes in chaotic transient dynamics,   we believe that this method can be extended, under minor modifications, to other interesting scenarios.

\enlargethispage{20pt}

\begin{acknowledgments}
This work was supported by the Spanish State Research Agency (AEI) and the European Regional Development Fund (FEDER) under Project No. FIS2016-76883-P.
\end{acknowledgments}

\end{document}